\documentclass[a4paper,11pt]{article}

\usepackage[utf8]{inputenc}
\usepackage[english]{babel} 
\usepackage[T1]{fontenc}
\usepackage{hyperref}	
\usepackage[acronym, nomain]{glossaries} 
\usepackage{amsthm}
\usepackage{amssymb}
\usepackage{mathrsfs}
\usepackage{parskip}		
\usepackage{txfonts}			
\usepackage{enumitem}			
\usepackage{algorithm}
\usepackage{algpseudocode} 
\usepackage{lipsum} 
\usepackage{tabularx}
\usepackage{stmaryrd}
\usepackage{float}
\usepackage{csquotes}
\usepackage{layout}
\usepackage{textcomp} 
\usepackage{authblk}
\usepackage{comgipp}

\usepackage[dvipsnames]{xcolor}
\usepackage{lstomdoc}
\usepackage{sectsty}
\subsectionfont{\normalfont\large\itshape}

\usepackage[a4paper]{geometry}
\usepackage{fancyhdr}

\usepackage{graphicx}
\usepackage{subfig} 
\usepackage{wrapfig}	
\graphicspath{ {./images/} } 

\usepackage[export]{adjustbox}  

\usepackage{cleveref}			

\usepackage[
backend		= biber,
style		= authoryear,
giveninits	= true, 	
uniquename	= init, 	
natbib		= true,
url			= false,
isbn			= false,
doi			= false,
eprint		= false,
maxbibnames	= 99,
minbibnames	= 3
]{biblatex}

\DeclareNameAlias{sortname}{last-first}
\newcommand{\theReferenceText}{\fullcite{GreinerPetter2019}}
\hypersetup{
	pdftitle	= {Semantic Preserving Bijective Mappings for Expressions involving Special Functions in Computer Algebra Systems and Document Preparation Systems}
}

\addto\captionsenglish{
	}

\apptocmd{\UrlBreaks}{\do\f\do\m}{}{}
\newcommand{\NULL}{\textbf{null}}
\renewcommand{\Return}{\textbf{return}}

\newcommand{\DLMF}{DLMF}
\newcommand{\DRMF}{DRMF}

\newcommand{\Maple}{Maple}
\newcommand{\Mathematica}{Mathematica}
\newcommand{\Macro}{\DLMF/\DRMF{} \LaTeX{} macro}

\newcommand{\JOBAD}{{\tt JOBAD}}

\newcommand{\tbs}{\textbackslash}

\newcommand{\langMaple}{\mathfrak{M}_{aple}}

\newcommand{\inertF}{\texttt{InertForm}}

\newcommand{\sTeX}{{\raisebox{-.5ex}S\kern-.5ex\TeX}}

\newcommand{\tableRowSpace}{\rule{0pt}{0.9\normalbaselineskip}}

\newcommand{\tempEnvCitation}{}
\newcommand{\tempEnvSpacing}{}

\newcommand*{\myRuleTextFill}[2]{  \makebox[#1]{    \leaders\hrule height \dimexpr 8.2pt\relax depth \dimexpr -7pt\relax \hfill    \enskip{#2}\enskip    \leaders\hrule height \dimexpr 8.2pt\relax depth \dimexpr -7pt\relax \hfill\kern0pt}}

\def\myRuleFill{\leavevmode\leaders\hrule height \dimexpr 8.2pt\relax depth \dimexpr -7pt\hfill\kern0pt}

\newtheoremstyle{defTheoStyle}
{6pt} 	{2pt} 	{\itshape} 	{} 	{\bfseries} 	{} 	{\newline} 	{\thmname{#1} \thmnumber{#2}: {\normalfont\textit{(#3)}}} 

\newtheoremstyle{defExampStyle}
{6pt} 	{2pt} 	{\itshape} 	{} 	{\bfseries} 	{} 	{\newline} 	{\thmname{#1} \thmnumber{#2}:} 

\theoremstyle{defTheoStyle}

\theoremstyle{defExampStyle}

\newacronym{dlmf}{DLMF}{Digital Library of Mathematical Functions}
\newacronym{drmf}{DRMF}{Digital Repository of Mathematical Formulae}
\newacronym{cas}{CAS}{Computer Algebra Systems}
\newacronym{mlp}{MLP}{Mathematical Language Parser}
\newacronym{mlp-pt}{PPT}{PoM-Parsed Tree}
\newacronym{pom-pt}{PPT}{PoM-Parsed Tree}
\newacronym{bnf}{BNF}{Backus-Naur Form}
\newacronym{p2c}{P2C}{Presentation-To-Computation}
\newacronym{nlp}{NLP}{Natural Language Processing}
\newacronym{pom}{PoM}{Part-of-Math}
\newacronym{csv}{CSV}{Comma-Separated Values}
\newacronym{json}{JSON}{JavaScript Object Notation}
\newacronym{teo}{TEO}{Translated Expression Object}
\newacronym{pol}{NPN}{Normal Polish Notation}
\newacronym{rpol}{RPN}{Reverse Polish Notation}
\newacronym{dag}{DAG}{Directed Acyclic Graph}
\newacronym{api}{API}{Application Programming Interface}
\newacronym{mathml}{MathML}{Mathematical Markup Language}
\newacronym{oop}{OOP}{Object-Oriented Programming}
\newacronym{cdf}{CDF}{Computable Document Format}
\newacronym{mathML}{MathML}{Mathematical Markup Language}
\newacronym{cfsf}{CFSF}{Continued Fractions for Special Functions}
\newacronym{ecf}{eCF}{Encoding Continued Fraction Knowledge}
\newacronym{nist}{NIST}{National Institute of Standards and Technology}
\newacronym{vmext}{VMEXT}{Visualizing Mathematical Expression Trees}
\newacronym{opsf}{OPSF}{Orthogonal Polynomials and Special Functions}
\newacronym{dps}{DPS}{Document Preparation Systems}
\newacronym{stem}{STEM}{Science, Technology, Engineering and Mathematics}
\newacronym{gui}{GUI}{Graphical User Interface}
\newacronym{omdoc}{OMDoc}{Open Mathematical Documents}

\DeclareRobustCommand{\cpi}{{\pi}}
\DeclareRobustCommand{\expe}{{e}}

\DeclareRobustCommand{\iunit}{{i}}

\DeclareRobustCommand{\Real}{\mathbb{R}}

\newcommand{\Gudermannian}[1]{\mathrm{gd}\!\left(#1\right)}
\newcommand{\atan}[1]{\mathrm{arctan}\!\left(#1\right)}

\newcommand{\ph}[1]{\mathrm{ph}\!\left(#1\right)}

\title{
	Semantic Preserving Bijective Mappings for Expressions involving Special Functions between Computer Algebra Systems and Document Preparation Systems
}

\author{Andr\'e Greiner-Petter\textsuperscript{1}}
\author{Moritz Schubotz\textsuperscript{1}}
\author{Howard S.~Cohl\textsuperscript{2}}
\author{Bela Gipp\textsuperscript{1}}

\affil{
	\textsuperscript{1}Information Science Group, University of Konstanz, Germany\\
	\url{{first.last}@uni-konstanz.de}
}

\affil{
	\textsuperscript{2}Applied and Computational Mathematics Division,\\
	National Institute of Standards and Technology, Mission Viejo, CA, USA,\\
	\url{howard.cohl@nist.gov}
}

\date{} 

\DeclareLabeldate[online]{%
	\field{date}
	\field{year}
	\field{eventdate}
	\field{origdate}
	\field{urldate}
}

\begin{document}
	\maketitle
	\thispagestyle{firststyle}
	\begin{abstract}
		\glsresetall
		\noindent
		\textbf{Purpose:} Modern mathematicians and scientists of math-related disciplines often use \gls*{dps} to write and \gls*{cas} to calculate mathematical expressions. Usually, they translate the expressions manually between \gls*{dps} and \gls*{cas}. This process is time-consuming and error-prone. Our goal is to automate this translation. This paper uses \Maple{} and \Mathematica{} as the \gls*{cas}, and \LaTeX{} as our \gls*{dps}.
		
		\noindent\textbf{Approach:} Bruce Miller at the \gls*{nist} developed a collection of special \LaTeX{} macros that create links from mathematical symbols to their definitions in the \gls*{nist} \gls*{dlmf}. We are using these macros to perform rule-based translations between the formulae in the \gls*{dlmf} and \gls*{cas}. Moreover, we develop software to ease the creation of new rules and to discover inconsistencies.
		
		\noindent\textbf{Findings:} We created 396 mappings and translated 58.8\% of \gls*{dlmf} formulae (2,405 expressions) successfully between \Maple{} and \gls*{dlmf}. For a significant percentage, the special function definitions in \Maple{} and the \gls*{dlmf} were different. Therefore, an atomic symbol in one system maps to a composite expression in the other system. The translator was also successfully used for automatic verification of mathematical online compendia and \gls*{cas}. Our evaluation techniques discovered two errors in the \gls*{dlmf} and one defect in \Maple.
		
		\noindent\textbf{Originality:} This paper introduces the first translation tool for special functions between \LaTeX{} and \gls*{cas}. The approach improves error-prone manual translations and can be used to verify mathematical online compendia and \gls*{cas}.
	\end{abstract} \glsresetall
	

	\section{Introduction}
	A typical workflow of a scientist who writes a scientific publication is to use \gls*{dps} to write the paper and one or more \gls*{cas} for verification, analysis and visualization. Especially in the \gls*{stem} literature, \LaTeX{}
	has become the de facto standard for writing scientific publications over the past 30 years~\parencites{Knuth}[559]{DigitalTypo}{LATEX:Standard}. \LaTeX{} enables printing of mathematical formulae in a structure similar to handwritten style. For example, consider the specific Jacobi polynomial~\parencite[Table 18.3.1]{NIST:DLMF}
	\begin{equation}\label{eq:P}
	P_n^{(\alpha , \beta)}(\cos(a\Theta)),
	\end{equation}
	where $n$ is a nonnegative integer, $\alpha, \beta > -1$, and $a,\Theta\in \Real$.
	This mathematical expression can be written in \LaTeX{} as
	\begin{equation*}
		\verb|P_n^{(\alpha,\beta)}(\cos(a\Theta))|.
	\end{equation*}
	
	While \LaTeX{} focuses on displaying mathematics, a \gls*{cas} concentrates on computations and user friendly syntax. Especially important for a CAS is to embed unambiguous semantic information within the input. Therefore, each system uses different representations and syntax, so that a writer needs to continually translate mathematical expressions from one representation to another and back again. Table~\ref{tab:JacobiP-usecase} shows four different representations for~(\ref{eq:P}).
	
	\begin{table}[ht]
		\centering
		\begin{tabular}{ll}
			\hline
			Systems & Representations \\
			\hline
			\hline
			Generic \LaTeX\ & \verb|P_n^{(\alpha,\beta)}(\cos(a\Theta))| \\ 
			Semantic \LaTeX\ & \verb|\JacobiP{\alpha}{\beta}{n}@{\cos@{a\Theta}}| \\
			\Maple & \verb|JacobiP(n,alpha,beta,cos(a*Theta))| \\ 
			\Mathematica & \verb|JacobiP[n,\[Alpha],\[Beta],Cos[a \[CapitalTheta]]]|\\
			\hline
		\end{tabular}
		\caption{Different representations for~\eqref{eq:P}. Generic \LaTeX{} is the default \LaTeX{} expression; semantic \LaTeX{} uses special semantic macros to embed semantic information; and \gls*{cas} representations are unique to themselves.}
		\label{tab:JacobiP-usecase}
	\end{table}
	
	Translations from generic \LaTeX{} to \gls*{cas} are difficult to realize since the full semantic information is not easily constructed from the input. Bruce Miller at the \gls*{nist} has created a set of semantic \LaTeX{} macros~\parencite{DLMF:Macros}. Each macro ties specific character sequences to a well-defined mathematical object and is linked with the corresponding definition in the \gls*{dlmf}. The \gls*{drmf} is an outgrowth of the \gls*{dlmf} with the goal to facilitate interaction among a community of mathematicians and scientists~\parencites{DRMF:14}{DRMF:15}. The \gls*{drmf} extends the set of semantic macros. These macros embed necessary semantic information into \LaTeX{} expressions. The macros may also contain $@$ symbols preceding the variables of the function. The number of $@$ symbols is used to switch between different notation styles, e.g., $\cos(x)$ and $\cos x$. One example of such a macro is given in Table~\ref{tab:JacobiP-usecase} for the semantic \LaTeX{} representation for the Jacobi polynomial. The macros provide isolated access to important parts of the mathematical function, such as the arguments. 
	
\begin{figure}[ht]
		\centering
		\subfloat[The real part of arccotangent with a branch cut at ${[-\iunit\infty, -\iunit]}$, ${[\iunit,\iunit\infty]}$.]{
			\includegraphics[width=0.45\textwidth]{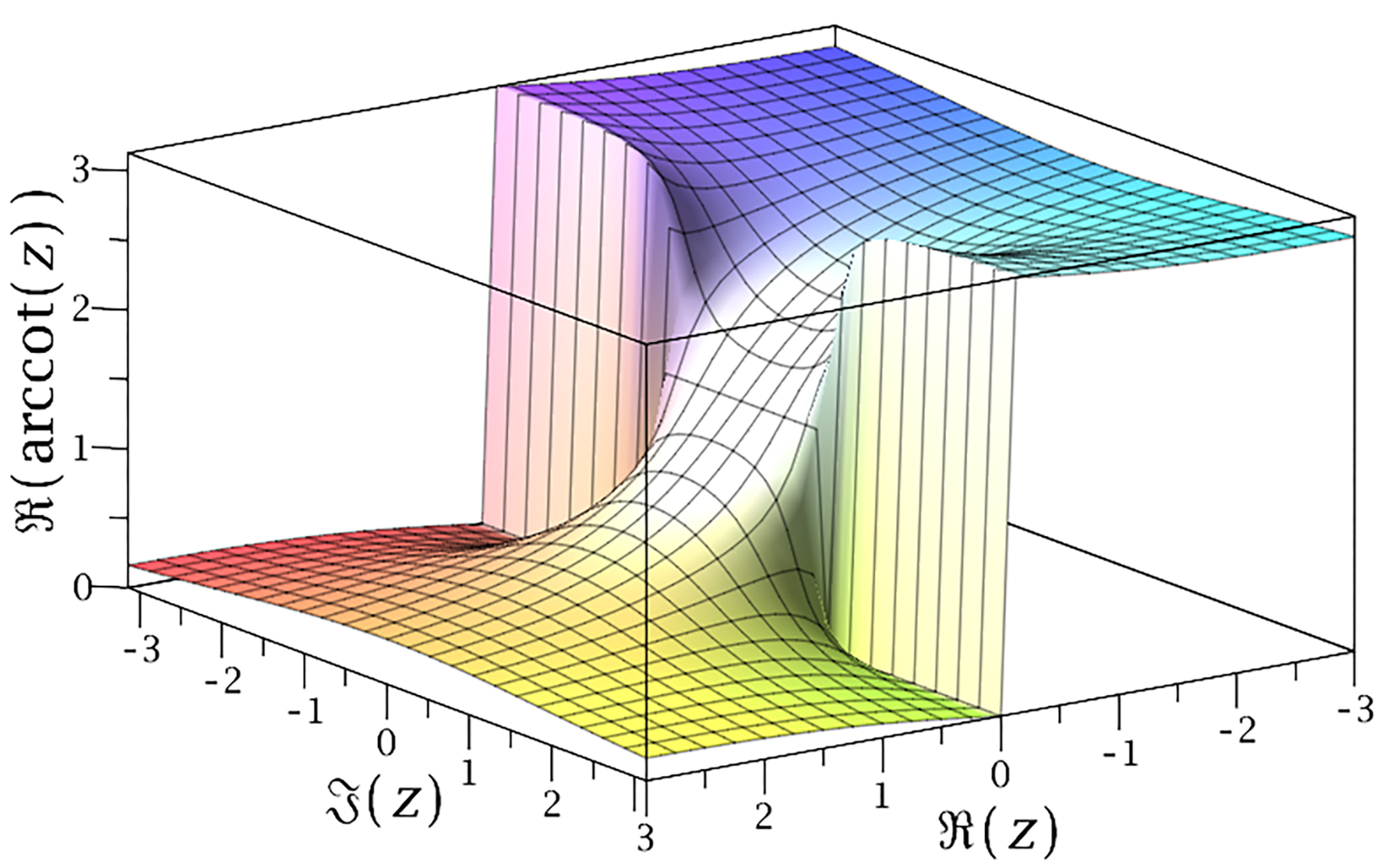}
			\label{fig:acot-cut1}
		}
		\hspace{0.5cm}
		\subfloat[The real part of arccotangent with a branch cut at ${[-\iunit, \iunit]}$.]{
			\includegraphics[width=0.45\textwidth]{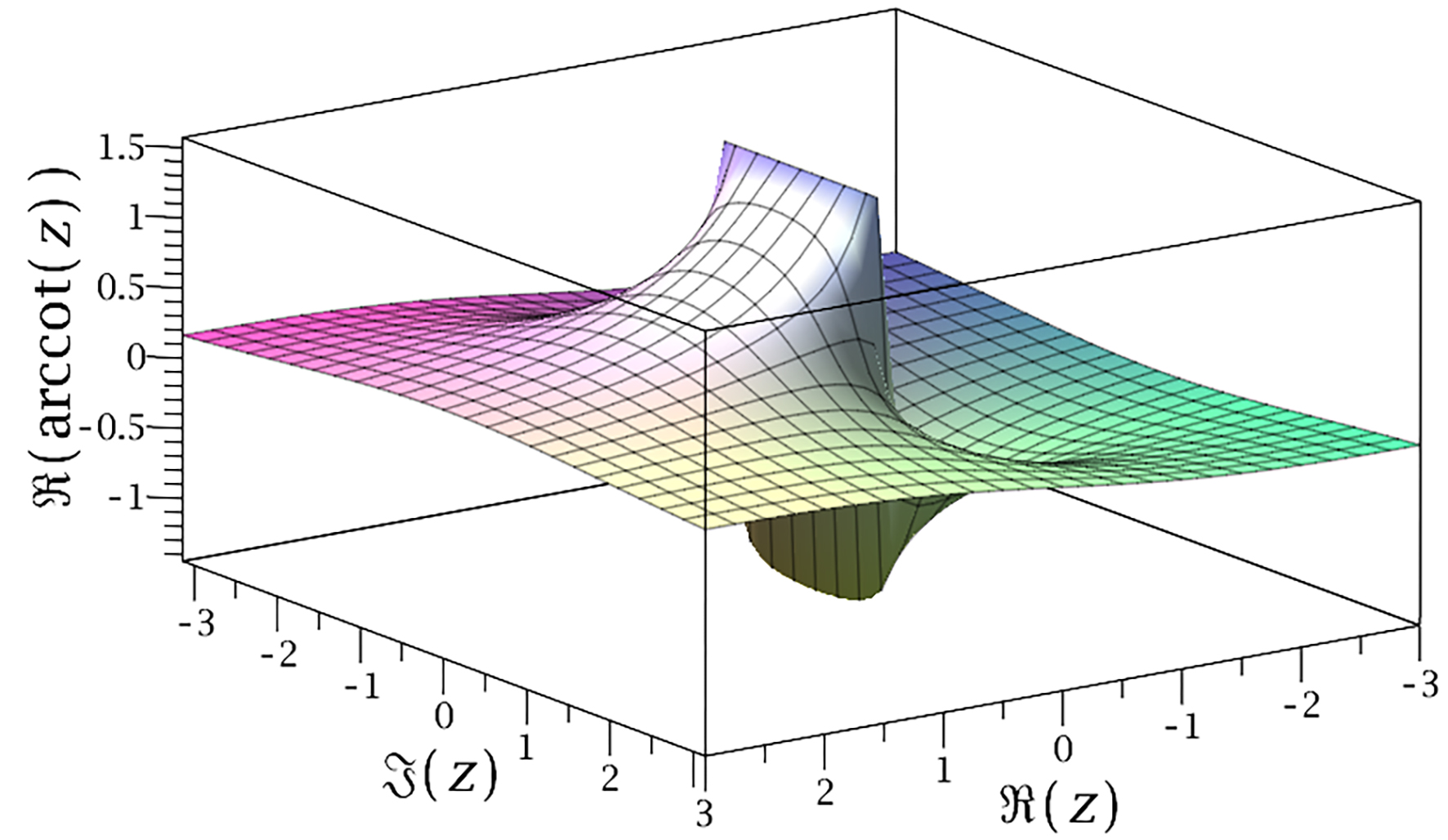}
			\label{fig:acot-cut2}
		}
		\caption{Two plots of the real part for the arccotangent function with a branch cut at $[-\iunit\infty, -\iunit]$, $[\iunit,\iunit\infty]$ in Figure~\protect\subref{fig:acot-cut1} and at $[-\iunit, \iunit]$ in Figure~\protect\subref{fig:acot-cut2}, respectively. (Plotted with \Maple{} 2016)}
		\label{fig:acot-cut-compare}
	\end{figure}

	Even with embedded semantic information, a translation between systems can be difficult. A typical example of complex problems occurs for multivalued functions~\parencite{AISC:MultivaluedFunctions}. A \gls*{cas} usually defines \textit{branch cuts} to compute principal values of multivalued functions~\parencite{Maple:Cuts}, which makes the implementation of a theoretically continuous function to a discontinuous presentation of it. In general, positioning branch cuts follows conventions, but can be positioned arbitrarily in many cases. Communicating and explaining the decision of defined branch cuts is a critical issue for \gls*{cas} and can vary between various systems~\parencite{Branches:acot}. Figure~\ref{fig:acot-cut-compare} illustrates two examples of different branch cut positioning for the inverse trigonometric arccotangent function. While \Maple{}\footnote{The mention of
		specific products, trademarks, or brand
		names is for purposes of identification only. Such mention is not to be interpreted
		in any way as an endorsement or certification of such products or brands by the
		National Institute of Standards and Technology, nor does it imply that the products
		so identified are necessarily the best available for the purpose. All trademarks
		mentioned herein belong to their respective owners.} defines the branch cut at ${[-\iunit\infty, -\iunit]}$, ${[\iunit,\iunit\infty]}$ (Figure~\ref{fig:acot-cut1}), \Mathematica{} defines the branch cut at ${[-\iunit, \iunit]}$ (Figure~\ref{fig:acot-cut2}).
	
	A \gls*{cas} user needs to fully understand the properties and special definitions (such as the position of branch cuts) in the \gls*{cas} to avoid mistakes during a translation~\parencite{Maple:Cuts}. A manual translation process is not only laborious, but also prone to errors. Note that this general problem has been named as automatic \gls*{p2c} conversion~\parencite{POM-Tagger}.
	
	This article presents a new approach for automatic \gls*{p2c} and vice versa conversions. Translations from presentational to computational (computational to presentational) systems are called forward (backward) translations. A forward translation is denoted with an arrow with the target system language above the arrow. For example,
	\begin{equation*}
		t \overset{\langMaple}{\mapsto} c,
	\end{equation*}
	where $t$ is an expression in the \LaTeX{} language and $c$ is an element of the \Maple{} language $\langMaple$. As we will see later in this article, we need to compare mathematical concepts between systems. This is impossible from a mathematical point of view. Consider the irrational mathematical constant $e$, known as Euler's number. The theoretical construct for this symbol cannot be mathematically equivalent to the value \verb|exp(1)| in \Maple, caused by computational and implementational limitations. 
	
	In order to clarify the notion of \textit{equivalence} (or lack thereof) in our context of translations,
	we introduce the terms \textit{appropriate} and \textit{inappropriate} translations. 
	We consider a translation to be {\it appropriate}, when a numerical evaluation returns the 
	same values in both concepts up to a numerical precision $|\epsilon|\ll 1$, for all possible 
	points in specified domains for the functions.
	A translation is considered as {\it inappropriate}, when it is not {\it appropriate}.
	
	For example, a translation such as
	\begin{equation}\label{eq:cos-def}
	\verb|\cos@{z}| \overset{\langMaple}{\mapsto} \verb|cos(z)|
	\end{equation}
	is \textit{appropriate}, while a translation such as
	\begin{equation}
	\verb|\cos@{z}| \overset{\langMaple}{\mapsto} \verb|sin(z)|
	\end{equation}
	is \textit{inappropriate}. Note that it is not always as easy as in this example to decide if a translation is {\it appropriate} or not. Therefore, this article also presents several validation techniques to automatically verify if a translation is {\it appropriate} or {\it inappropriate}. 
	
	In addition, we also introduce the notion of \textit{direct} translations. Most mathematical objects in one system have a direct counterpart in other systems. Later in the paper, we will explain that a translation from one specific mathematical object to its counterpart in the other system is not always \textit{appropriate}. Also, not every mathematical object has a counterpart in other systems. We call a translation to its counterpart \textit{direct}. For example, the translation~(\ref{eq:cos-def}) is \textit{direct}, while a translation to the definition of the cosine function
	\begin{equation*}
		\verb|\cos@{z}| \overset{\langMaple}{\mapsto} \verb|(exp(I*z)+exp(-I*z))/2|
	\end{equation*}
	is not a \textit{direct} translation even though it is {\it appropriate}.
	Note that partial results of this paper have been published in~\parencite{CICM:Paper}.
	
	\section{Related Work}\label{sec:related-work}
	Since \LaTeX{} became the de facto standard for writing papers in mathematics, most \gls*{cas} provide simple functions to import and export mathematical \LaTeX{} expressions\footnote{The selected \gls*{cas} Maple, Mathematica, Matlab, and SageMath provide import and/or export functions for \LaTeX:\\
		Maple, \url{http://www.maplesoft.com/support/help/Maple/view.aspx?path=latex} seen 06/2017;
		Mathematica, \url{https://reference.wolfram.com/language/tutorial/GeneratingAndImportingTeX.html} seen 06/2017;
		Matlab, \url{https://www.mathworks.com/help/symbolic/latex.html} seen 06/2017;
		SageMath, \url{http://doc.sagemath.org/html/en/tutorial/latex.html} seen 06/2017.}. Those tools have two essential problems. They are only able to import simple mathematical expressions, where the semantics are unique. For example, the internal \LaTeX{} macro \verb|\frac| always indicates a fraction. For more complex expressions, e.g., the Jacobi polynomial in \Cref{tab:JacobiP-usecase}, the import functions fail. The second problem appears in the export tools. Mathematical expressions in \gls*{cas} are fully semantic. Otherwise the \gls*{cas} wouldn't be able to compute or evaluate the expressions. During the export process, the semantic information is lost, because generic \LaTeX{} is not able to carry sufficient semantic information. Because of these problems, an exported expression cannot be imported to the same system again in most cases (except for simple expressions such as those described above). Our tool attempts to solve these problems and provide round-trip translations between \LaTeX{} and \gls*{cas}.
	
	The semantics must be well-known before an expression can be translated. There are two main approaches to solve that problem:~(1) someone could specify the semantic information during the writing process (pre-defined semantics); and (2) the translator can determine the correct semantic information in general mathematical expressions before it translates the expression. So-called \textit{interactive documents}\footnote{There is no adequate definition for what interactive documents are. However, this name is widely used to describe electronic document formats that allow for interactivity to change the content in real time.}, such as the \gls*{cdf}\footnote{Wolfram Research; \textit{Computable Document Format} (CDF); \url{http://www.wolfram.com/cdf/}, July 2011} by Wolfram Research, or \textit{worksheets} by \Maple{}, try to solve this problem with the approach (2) and allow one to embed semantic information into the input. Those complex document formats require specialized tools to show and work with the documents (Wolfram CDF Player, or \Maple{} for the \textit{worksheets}). The \JOBAD{} architecture~\parencite{JOBAD:orig} is able to create web-based interactive documents and uses \gls*{omdoc}~\parencite{OMDoc} to carry semantics. The documents can be viewed and edited in the browser. Those \JOBAD{-documents} also allow one to perform computations via \gls*{cas}. This gives one the opportunity to calculate, compute and change mathematical expressions directly in the document. The translation performs in the background, invisible to the user. Similar to the \JOBAD{} architecture, other interactive web documents exist, such as \textit{MathDox}~\parencite{MathDox} and \textit{The Planetary System}~\parencite{Planetary}.
	
	Another approach tries to avoid translation problems by allowing computations directly via the \LaTeX{} compiler, e.g., \textit{LaTeXCalc}~\parencite{LatexCalc}. Those packages are limited to the abilities of the compiler and therefore are not as powerful as \gls*{cas}. A workaround for this case is \textit{sagetex}~\parencite{Sagetex}, which is a \LaTeX{} package interface for the open source \gls*{cas} \textit{sage}\footnote{An abbreviation for \textit{SageMath}.}. This package allows \textit{sage} commands in \TeX{-}files and uses \textit{sage} in the background to compute the commands. In this scenario, a writer still needs to manually translate expressions to the syntax of \textit{sage}, but it is possible to integrate \gls*{cas} expressions directly into \TeX{} documents.
	
	There exist two approaches for marking up mathematical \TeX/\LaTeX{} documents semantically with \TeX{} macros. Namely, \sTeX{}~\parencite{sTeX} developed by Kohlhase and the \Macro s developed by Miller~\parencite{DLMF:Macros}. This paper shows that it is possible to develop a context-free translation tool using  the semantic macros introduced by Miller. The goal of \sTeX{} is to markup the functional structure of mathematical documents so that they can be exported to the \gls*{omdoc} format. The macro functionality developed by Miller introduces new macros for special functions, orthogonal polynomials, and mathematical constants. Each of these macros ties specific character sequences to a well-defined mathematical object and is linked with the corresponding definition in the \gls*{dlmf} or \gls*{drmf}. Therefore, we call these semantic macros \Macro s. These semantic macros are internally used in the \gls*{dlmf} and the \gls*{drmf}. We gave the \Macro{} set the favor for developing the translation engine because it provides \gls*{dlmf} definitions for a comprehensive number of functions. In contrast, \sTeX{} does not focus on the semantics of functions, is often complex to use, and defines diverse macros for symbols and concepts that \gls*{cas} usually does not usually support.
	
	Miller also developed LaTeXML, a tool for converting \LaTeX{} expressions to MathML~\parencite{LaTeXML}. LaTeXML is used to generate the \gls*{dlmf} and is able to parse the \Macro s to generate content MathML. Even though many \gls*{cas} are able to import and export MathML, they fail for special functions. Schubotz and collaborators recently performed benchmarks on several \LaTeX{} to MathML conversion tools, including LaTeXML, in~\parencite{DBLP:conf/jcdl/SchubotzGSMCG18}.
	
	\section{Translation Problems}\label{sec:problems}
	There are several potential problems for performing translations between systems that embed semantic information in the input. These problems vary from simple cases, e.g., a function is not defined in the system, to complex cases, e.g., different positioning of branch cuts for multivalued functions. This section will discuss the problems and our workarounds.
	
	\subsection{Different Sets of Defined Functions}
	If a function is defined in one system but not in the other, sometimes we can easily translate the definition of the mathematical function. For example, the \textit{Gudermannian}~\parencite[(4.23.10)]{NIST:DLMF} $\Gudermannian{x}$ function is defined by
	\begin{equation}\label{eq:gudermannian}
	\Gudermannian{x} := \atan{\sinh{x}}, \quad x \in \Real,
	\end{equation}
	and linked to the semantic macro \verb|\Gudermannian| in the \gls*{dlmf} but does not exist in \Maple. We can perform a translation for the definition~\eqref{eq:gudermannian} instead of macro itself
	\begin{equation}\label{eq:guder-trans}
	\verb|\Gudermannian{x}| \overset{\langMaple}{\mapsto} \texttt{arctan(sinh(x))}.
	\end{equation}
	
	\vspace{-0.2cm}
	Since translations such as these are nonintuitive, describing explanations become necessary for the translation process. A particular logging function stores each translation and provides details after a successful translation process. \Cref{sec:forward-translation} explains this task further.
	
	Providing detailed information also solves the problem for multiple alternative translations. In some cases, a semantic macro has two alternative representations in the \gls*{cas} or vice versa. In such cases, the translator picks one of the alternatives and informs the user about the decision.
	
	\subsection{Positions of Branch Cuts}
	In case of differences between defined branch cuts, we can also use alternative translations to solve the problems. Consider the mentioned case of the arccotangent function~\parencite{Branches:acot} that has different positioned branch cuts in \Maple{} as compared to the \gls*{dlmf} or \Mathematica{} definitions. As suggested by~\parencite{Branches:acot}, we can translate an alternative definition of the arccotangent function to avoid the branch cut issues. Considering 
	\parencite[(23), (25)]{Branches:acot}, we can define three translations
	\begin{eqnarray}
	\verb|\acot@{z}| & \overset{\langMaple}{\mapsto} & \verb|arccot(z)|,\label{eq:acot-alternatives}\\
	& \overset{\langMaple}{\mapsto} & \verb|arctan(1/z)|,\label{eq:acot-alternatives-1}\\
	& \overset{\langMaple}{\mapsto} & \verb|I/2*ln((z-I)/(z+I))|.\label{eq:acot-alternatives-2}
	\end{eqnarray}
	
	\vspace{-0.2cm}
	The position of the branch cut of the arccotangent function differs after the {\it direct} translation~\eqref{eq:acot-alternatives}, which may lead to incorrect calculations later on. The alternative translations~\eqref{eq:acot-alternatives-1} and~\eqref{eq:acot-alternatives-2} use other functions instead of the arccotangent function. The arctangent function~\eqref{eq:acot-alternatives-1} and the natural logarithm~(\ref{eq:acot-alternatives-2}) have the same positioned branch cuts as in the \gls*{dlmf} and in \Maple. Therefore, translation~(\ref{eq:acot-alternatives-1}) solves this issue as long as the user does not evaulate the function at $z = 0$, while translation~(\ref{eq:acot-alternatives-2}) solves the issue except at $z=-\iunit$. Note that none of the translations~(\ref{eq:acot-alternatives}-\ref{eq:acot-alternatives-2}) are {\it appropriate}.
	
	\subsection{Insufficient Semantic Information}
	Other problematic cases for translations are the \Macro s themselves. In some cases, they do not provide sufficient semantic information to perform translations. One example is the \textit{Wronskian} determinant. For two differentiable functions $w_1$, $w_2$, the \textit{Wronskian} is defined as~\parencite[(1.13.4)]{NIST:DLMF}
	\begin{equation*}
		\mathscr{W}\{ w_1(z), w_2(z) \} = w_1(z)w_2'(z) - w_2(z)w_1'(z).
	\end{equation*}
	In semantic \LaTeX{}, it is currently implemented using
	\begin{equation}
	\verb|\Wronskian@{w_1(z), w_2(z)}|.
	\end{equation}
	This translation is unfeasible because the macro does not explicitly define the variable of differentiation for the functions $w_1$, $w_2$. For a correct translation, the \gls*{cas} needs to be aware of the variable of differentiation $z$. We solved this issue by creating a new macro \verb|\Wron|, e.g.,
	\begin{equation}
	\verb|\Wron{z}@{w_1(z)}{w_2(z)}|.
	\end{equation}
	This example demonstrates that the \Macro s are still a work in progress and further updates are sometimes
	necessary in order to further encapsulate critical semantic information.
	
	\subsection{Potentially Ambiguous Expressions}
	Since the \Macro s aims to cover an extensive set of special functions, orthogonal polynomials, and mathematical constants, they does not contain specific macros for other mathematical objects. However, also mathematical expression without functions, polynomials and mathematical constants can be ambigious. As an example, multiplications are rarely explicitly marked in \LaTeX{} expressions, e.g., scientists using whitespace to indicate multiplications rather than using \verb|\cdot| or similar symbols. But whitespaces can also be used to improve the readability and not to represent a multiplication.
	
	For such problems, we introduced a new macro \verb|\idot| for an invisible multiplication symbol (this macro will not be rendered). Since this macro is newly introduced by contributers of the \gls*{drmf} team, and automatic conversion of existing equations is difficult, none of the equations in the \gls*{dlmf} use this macro. Therefore, the translator has some simple 
	rules for performing translations without explicitly marking multiplication translations with \verb|\idot|.
	
	The \Macro s do not guarantee entirely disambiguated expressions. In~\Cref{tab:amb-latex} there are four examples of potentially ambiguous expressions. These expressions are unambiguous for the \LaTeX{} compiler since it only considers the very next token for superscripts and subscripts. Our translator follows the same rules to solve these issues.
	
	\begin{table}[ht]
		\centering
		\begin{tabular}{cc}
			\hline
			Potentially Ambiguous Input & \LaTeX{} Output\\
			\hline
			\verb|n^m!| & $n^m!$\\
			\verb|a^bc^d| & $a^bc^d$\\
			\verb|x^y^z| & Double superscript error\\
			\verb|x_y_z| & Double subscript error\\
			\hline
		\end{tabular}
		\caption{Potentially ambiguous \LaTeX{} expressions and how \LaTeX{} displays them.}
		\label{tab:amb-latex}
	\end{table}
	
	Another more questionable translation decision addresses alphanumerical expressions. As explained in~\Cref{tab:allTypesTable}, the \gls*{pom}-tagger handles strings of letters and numbers differently depending on the order of the symbols. The reason is that an expression such as `$4b$' is usually considered to be a multiplication of $4$ and `$b$,' while `$b4$' gives the impression that $4$ indexing `$b$'. While the first example produces two nodes, namely $4$ and `$b$', the second example `$b4$' produces just a single alphanumerical node in the \gls*{pom-pt}. The translator interprets alphanumerical expressions as multiplications for two reasons:~(1) we would assume that the inputs `$4b$' and `$b4$' are mathematically equivalent; and (2) it is more common in mathematics to use single letter names for variables~\parencite{Notation:History}. Therefore we have used rules as follows
	
	\begin{eqnarray*}
		\verb|4b| & \overset{\langMaple}{\mapsto} & \verb|4*b|,\\
		\verb|b4| & \overset{\langMaple}{\mapsto} & \verb|b*4|,\\
		\verb|energy| & \overset{\langMaple}{\mapsto} & \verb|e*n*e*r*g*y|.
	\end{eqnarray*}
	
	In general, the translator is designed to find a work-around for disambiguating expressions. 
	If there is no way to solve a potential ambiguity with defined rules, then we stop the translation process.
	
	\section{The Translator}
	The translator analyzes a parse tree to perform translations. For generating a parse tree of \LaTeX{} expressions, the translator uses the \gls*{pom}-Tagger~\parencite{POM-Tagger}\footnote{Named according to the Part-of-Speech-Taggers in \gls*{nlp}.}. \gls*{cas} define their own syntax parser. We were able to use \Maple's internal data structure to 
	obtain a parse tree of the input. \Cref{sec:forward-translation} and \Cref{sec:backward-translation} will explain the parsing and translation process in detail.
	
	All translations are defined by a library (\gls*{csv} and \gls*{json} files) that define translation patterns for each function and symbol. The pattern uses \verb|$i| as a placeholder to determine the positions of the arguments. For example, the translation patterns for the Jacobi polynomial are illustrated in~\Cref{tab:placeholder_ex2}.
	
	\begin{table}[ht]
		\centering
		\begin{tabular}{lc}
			\hline
			\multicolumn{2}{l}{\textit{Forward Translation:}} \\
			\Maple & \verb|JacobiP($2, $0, $1, $3)| \\
			\Mathematica & \verb|JacobiP[$2, $0, $1, $3]|\\
			\hline
			\multicolumn{2}{l}{\textit{Backward Translation from \Maple/\Mathematica:}} \\
			Semantic \LaTeX & \verb|\JacobiP{$1}{$2}{$0}@{$3}|\\
			\hline
		\end{tabular}
		\caption{Forward and backward translation patterns for the Jacobi polynomial example~\eqref{eq:P} in this manuscript. The pattern for the backward translation is the same for \Maple{} and \Mathematica.}
		\label{tab:placeholder_ex2}
	\end{table}
	
	The \Macro s also allow one to specify optional arguments to distinguish between standard and another version of these functions. The Legendre and associated Legendre functions of the first kind are examples of such cases. The library that defines translations for each macro uses the macro name as the primary key to identify the translations. The Legendre and associated Legendre function of the first kind both use the same macro \verb|\LegendreP|. To distinguish such cases, we use a special syntax, shown in~\Cref{tab:legendreP-lex}.
	
	\begin{table}[ht!]
		\centering
		\begin{tabular}{ll}
			\hline
			Semantic Macro Entry & \Maple{} Entry \\
			\hline
			\verb|\LegendreP{\nu}@{x}| & \verb|LegendreP($0, $1)| \\
			\verb|X1:\LegendrePX\LegendreP[\mu]{\nu}@{x}| & \verb|LegendreP($1, $0, $2)|\\
			\hline
		\end{tabular}
		\caption{Example entries of the Legendre and associated Legendre function in the translation library. The prefix notation \texttt{X<d>:<name>X} defines the translation for \texttt{<name>} with \texttt{<d>}-number of optional arguments.}
		\label{tab:legendreP-lex}
	\end{table}
	
	\subsection{Escape the Placeholder Symbol}
	The used placeholders cause trouble when the \gls*{cas} uses the symbol \verb|$| for other reasons, e.g., differentiation in \Maple{} is implemented as
	\begin{equation*}
		\verb|diff(f, [x$n])|,
	\end{equation*}
	where \verb|f| is an algebraic expression or an equation, \verb|x| is the name of the differentiation variable, 
	and \verb|n| is an integer representing the $n$-th order differentiation\footnote{\url{https://www.maplesoft.com/support/help/maple/view.aspx?path=diff}, seen 07/2018}. A translation for $\frac{d^2x^2}{dx^2}$ should display as
	\begin{equation*}
		\verb|\deriv[2]{x^2}{x}| \overset{\langMaple}{\mapsto} \verb|diff(x^2, [x$2])|
	\end{equation*}
	but would end up as
	\begin{equation*}
		\verb|\deriv[2]{x^2}{x}| \overset{\langMaple}{\mapsto} \verb|diff(x^2, [xx])|.
	\end{equation*}
	
	We can solve this issue by using parentheses in such cases, e.g., \verb|diff($1, [$2$($0)])|.
	
	\section{Forward Translations}\label{sec:forward-translation}
	As a pre-processing step, we use the \gls*{pom}-Tagger~\parencite{POM-Tagger}\footnote{Named according to the Part-of-Speech-Taggers in \gls*{nlp}.} for parsing semantic \LaTeX{} expressions. The \gls*{pom}-Tagger is defined by a context-free grammar in \gls*{bnf} and is an \texttt{LL}-Parser, i.e., it parses the input from \textbf{L}eft to right and assigns the \textbf{L}eftmost (first applicable) derivation rule defined by the grammar to an expression. In other words, the \gls*{pom}-Tagger scans the input for \textit{terms} and groups them into subexpressions if suitable, where \textit{terms} are non-terminal symbols in the context of \gls*{bnf}. A node in the generated parse tree will be tagged by meta information if the node matches defined symbols. The meta information is stored in lexicon files. Those lexicon files were manually cultivated for the \gls*{pom}-Tagger. 
	
	We integrated the defined translation patterns from our library also into these lexicon files. The tagger also tags a node in the parse tree by its translation patterns. \Cref{tab:sine-lex-example} gives an example of an entry of the lexicon file.
	
	\begin{table}[ht!]
		\centering
		\begin{tabular}{lll}
			\hline
			\multicolumn{3}{l}{Symbol:~\texttt{\textbackslash sin}} \\
			\! & \multicolumn{2}{l}{Feature Set:~dlmf-macro} \\
			\! & \! & DLMF:~\verb|\sin@@{z}|\\
			\! & \! & DLMF-Link:~dlmf.nist.gov/4.14.E1\\
			\! & \! & Meanings:~Sine\\
			\! & \! & Number of Parameters:~0\\
			\! & \! & Number of Variables:~1\\
			\! & \! & Number of Ats:~2\\
			\! & \! & Maple:~\verb|sin($0)|\\
			\! & \! & Maple-Link:~www.maplesoft.com/support/\\
			\! & \! & \hspace{32pt} help/maple/view.aspx?path=sin\\
			\! & \! & Mathematica:~\verb|Sin[$0]|\\
			\! & \! & Mathematica-Link:~reference.wolfram.com/\\
			\! & \! & \hspace{32pt} language/ref/Sin.html\\
			\hline
		\end{tabular}
		\caption{The entry of the trigonometric sine function in the lexicon file.}
		\label{tab:sine-lex-example}
	\end{table}
	
	The parsed tree generated by the \gls*{pom}-Tagger is not a mathematical expression tree. The \gls*{pom} project aims to disambiguate mathematical \LaTeX{} expressions and generates an expression tree. In the current state, however, many expressions still cannot be disambiguated. Therefore, the \gls*{pom}-tagger generates a raw parsed tree where each token in the \LaTeX{} expression is a node in the tree. We call this parsed tree the \gls*{pom-pt}.
	
\begin{figure}[ht]
		\vspace{-10pt}
		\centering
		\includegraphics[clip, trim=0.2cm 0.2cm 0.2cm 0.2cm, scale=0.72]{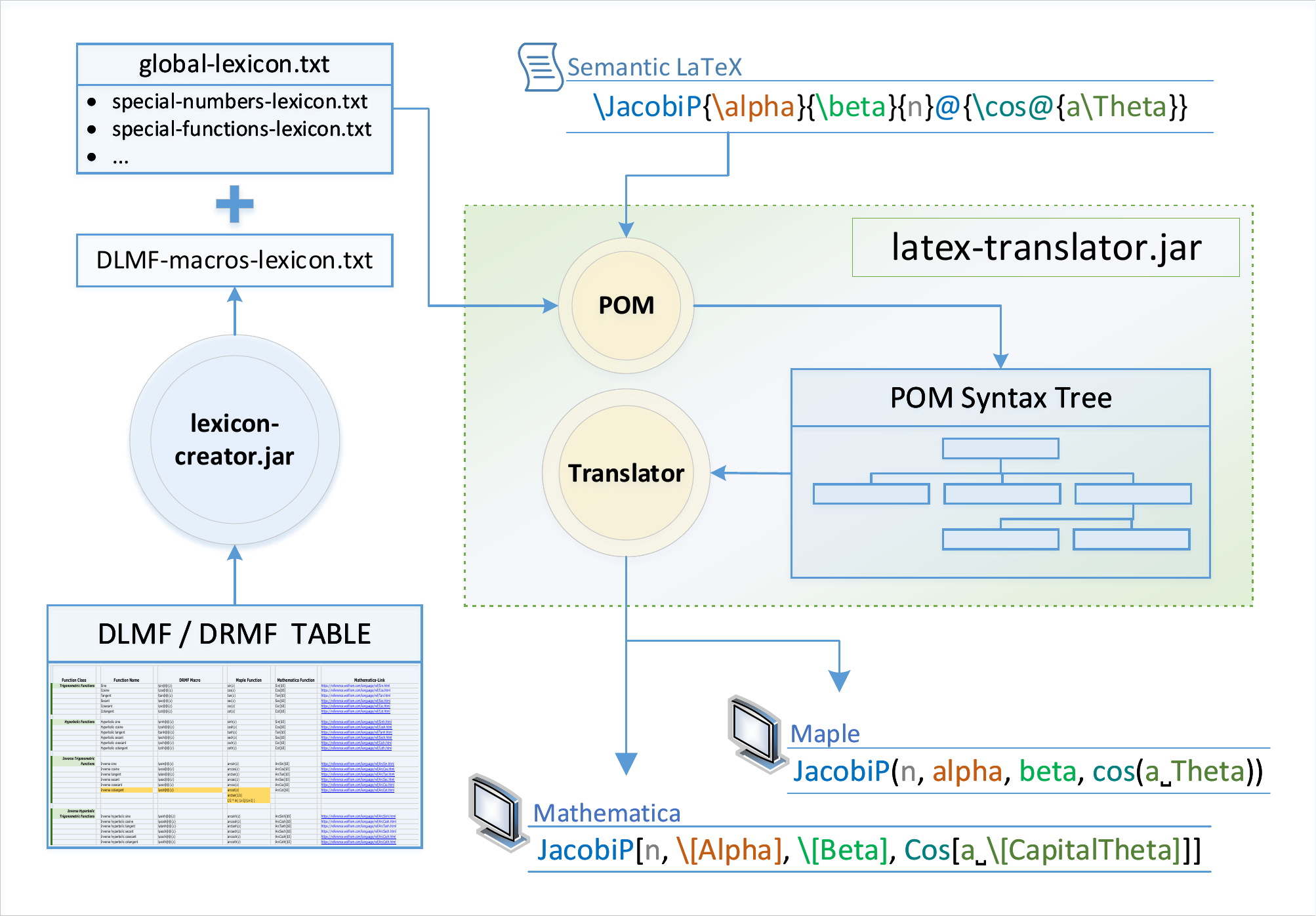}
		\caption{Process diagram of a forward translation process. The \gls*{pom}-tagger generates the \gls*{pom-pt} based on lexicon and \gls*{json} files. The \gls*{pom-pt} will be translated to different \gls*{cas}.}
		\label{fig:forward-trans}
		\vspace{-10pt}
	\end{figure}

	{\sloppy The overall forward translation process is explained in~\Cref{fig:forward-trans}. All translation patterns and related information are stored in the DLMF/DRMF tables. These tables are converted by the \verb|lexicon-creator| to the \verb|DLMF-macros-lexicon| lexicon file. Together with the \verb|global|-\verb|lexicon| file, the \gls*{pom-pt} will be created by the \gls*{pom}-tagger. The \verb|latex-converter| takes a string representation of a semantic \LaTeX{} expression and uses the \gls*{pom} engine as well as our \verb|Translator| to create a proper string representation for a specified \gls*{cas}.}
	
	\subsection{Analyzing the PoM-Parsed Tree}\label{subsec:analyze-mlp}
	Since the \gls*{bnf} does not define rules for semantic macros, each argument of the semantic macro and each $@$ symbol are following siblings of the semantic macro node. That is the reason why we stored the number of parameters, variables and $@$ symbols in the lexicon files. Otherwise, the translator could not find the end of a semantic macro in the \gls*{pom-pt}.
	
	\begin{figure}[ht]
		\centering
		\includegraphics[clip, trim=0.2cm 0.2cm 0.2cm 0.2cm, scale=0.75]{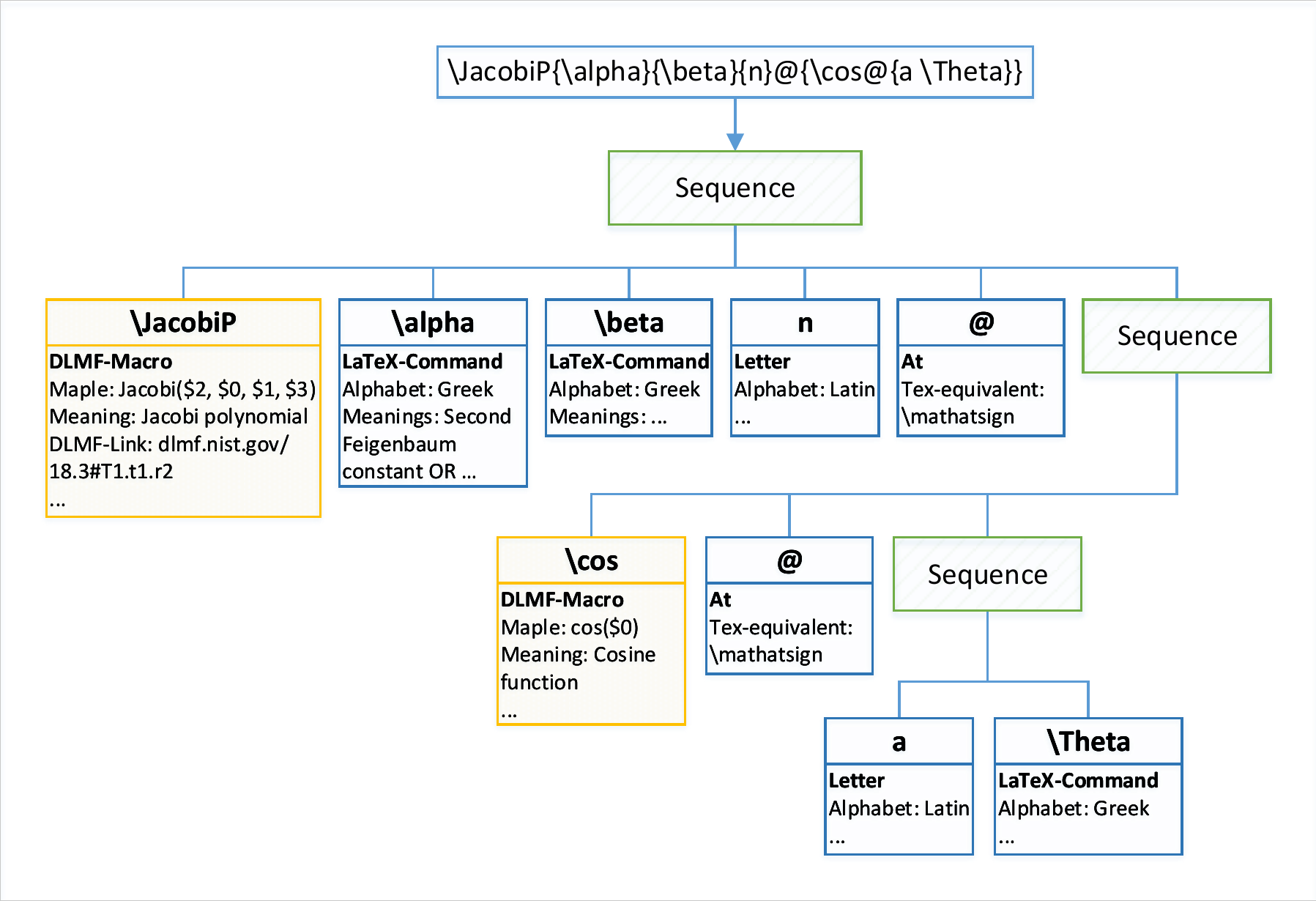}
		\caption{The \gls*{pom-pt} for the Jacobi polynomial example~\eqref{eq:P} using the DLMF/DRMF \LaTeX{} macro. Each leaf contains information from the lexicon files.}
		\label{fig:syntax-tree-usecase}
		\vspace{-10pt}
	\end{figure}
	
	\Cref{fig:syntax-tree-usecase} visualizes the \gls*{pom-pt} of the Jacobi polynomial example from~\Cref{tab:JacobiP-usecase}. Because of the differences between expression trees and \gls*{pom-pt}, it can be difficult to generate a string representation after a successful translation process. It is especially difficult to determine necessary and unnecessary parentheses when we analyse the \gls*{pom-pt}. Therefore we create the \gls*{teo}. The \gls*{teo} is a list containing already translated subexpressions.
	
	With these tools, we can translate a \LaTeX{} expression by translating the \gls*{pom-pt} node by node and perform group or reordering operations for some special cases. The algorithm is realized in a simple recursive structure. Whenever the algorithm finds a leaf, it can translate this single term. If the node is not a leaf, it starts to translate all children of the node recursively. This idea appears to be a practical and elegant solution, but it has a significant drawback. It cannot be used to translate functions. Since the arguments of functions are following siblings in the \gls*{pom-pt}, the algorithm needs to lookahead when a leaf is a known function, e.g., in the case of a semantic macro with arguments (see~\Cref{fig:syntax-tree-usecase}). Algorithm~\ref{alg:translation} is an improved version with lookahead functionality.
	
	\begin{algorithm}[t]
		\caption{Abstract translation algorithm to translate PPT.}\label{alg:translation}
		\begin{algorithmic}[1]
			\Require Root $r$ of a PoM-Parse tree $T$. List \textit{following\_siblings} with the following siblings of $r$. The list can be empty.
			\Procedure{abstract\_translator}{$r$, \textit{following\_siblings}}
			\If{$r$ is leaf}
			\State {\scriptsize TRANSLATE\_LEAF}($r$, \textit{following\_siblings});\label{line:transLeaf}
			\Else
			\State \textit{children} $ = r$.getChildren(); \Comment{\textit{children} is a list of nodes}
			\State {\scriptsize ABSTRACT\_TRANSLATOR}(\textit{children}.removeFirst(), \textit{children});\label{line:transSeq}
			\EndIf
			\If{\textit{following\_siblings} is not empty}\label{line:ifNotEmpty}
			\State $r =$ \textit{following\_siblings}.removeFirst();
			\State {\scriptsize ABSTRACT\_TRANSLATOR}($r$, \textit{following\_siblings});
			\EndIf
			\EndProcedure
		\end{algorithmic}
	\end{algorithm}
	
	If the root $r$ is a leaf, it still can be translated as a leaf. Eventually, some of the following siblings are needed to translate $r$. The list of \textit{following\_siblings} in Line~\ref{line:transLeaf} might be reduced to avoid multiple translations for one node. If $r$ is not a leaf, it contains one or more children. Therefore, we can call the {\footnotesize \verb|ABSTRACT_TRANSLATOR|} recursively for the children. Once we have translated $r$, we can go a step further and translate the next node. Line~\ref{line:ifNotEmpty} checks if there are following siblings left and calls the {\footnotesize \verb|ABSTRACT_TRANSLATOR|} recursively in such cases. Translated expressions are stored by the \gls*{teo} object. Algorithm~\ref{alg:translation} is a simplified version of the translator process. The \Cref{line:transLeaf,line:transSeq} process the translations for each node. \Cref{tab:allTypesTable} gives an overview of all the different node types the root $r$ can be. A more detailed explanation of the types can be found in~\parencite{POM-Tagger}.
	
	The \gls*{bnf} grammar defines some basic grammatical rules for generic \LaTeX{} macros, such as for \verb|\frac|, \verb|\sqrt|. Therefore, there is a hierarchical structure for those symbols similar to the structure in expression trees. As already mentioned, some of these types can be translated directly, such as Greek letters, while others are more complex, such as semantic \LaTeX{} macros. Therefore, the translators delegate the translation to specialized subtranslators. This delegation process is implemented in \Cref{line:transLeaf,line:transSeq} of Algorithm~\ref{alg:translation}.  
	Subsection~\ref{sec:subtranslators} discusses these classes in more detail.
	
	\begin{table}[t!]
		\centering
		\sloppy
		\newcolumntype{L}{>{\raggedright\arraybackslash}X}
		\newcolumntype{Y}[1]{>{\raggedright\let\newline\\\arraybackslash\hspace{0pt}}p{#1}}
		\begin{tabularx}{\textwidth}{Y{1.55cm} Y{2.5cm} L L}
			\hline
			& \textbf{Node type} & \textbf{Explanation} & \textbf{Example} \\
			\hline
			\hline
			\multicolumn{1}{Y{1.5cm}|}{\textbf{$r$ has children}} & Sequence & Contains a list of expressions. & $a+b$ is a sequence with three children ($a$, $+$ and $b$).\\
			\cline{2-4}
			\multicolumn{1}{l|}{} & Balanced Expression & Similar to a sequence. But in this case the sequence is wrapped by \texttt{\tbs left} and \texttt{\tbs right} delimiters. Note that normal parentheses do not create balanced expressions. & \texttt{\tbs left(}$a+b$\texttt{\tbs right)} is a balanced expression with three children ($a$, $+$ and $b$). \\
			\cline{2-4}
			\multicolumn{1}{l|}{} & Fraction & All kinds of fractions, such as \texttt{\tbs frac}, \texttt{\tbs ifrac}, etc. & \texttt{\tbs ifrac\{a\}\{b\}} is a fraction with two children ($a$ and $b$).\\
			\cline{2-4}
			\multicolumn{1}{l|}{} & Binomial & Binomials & \texttt{\tbs binom\{a\}\{b\}} has two children ($a$ and $b$).\\
			\cline{2-4}
			\multicolumn{1}{l|}{} & Square Root & The square root with one child. & \texttt{\tbs sqrt\{a\}} has one child ($a$).\\
			\cline{2-4}
			\multicolumn{1}{l|}{} & Radical with a specified index & $n$-th root with two children. & \texttt{\tbs sqrt[a]\{b\}} has two children ($a$ and $b$).\\
			\cline{2-4}
			\multicolumn{1}{l|}{} & Underscore & The underscore `\_' for subscripts. & The sequence $a$\_$b$ has two children ($a$ and `\_'). The underscore itself `\_' has one child ($b$).\\
			\cline{2-4}
			\multicolumn{1}{l|}{} & Caret & The caret `\^{}' for superscripts or exponents. Similar to the underscore. & The sequence $a$\^{}$b$ has two children ($a$ and `\^{}'). The caret itself `\^{}' has one child ($b$).\\
			\hline
			\hline
			\multicolumn{1}{Y{1.5cm}|}{\textbf{$r$ is a leaf}} & \Macro{} & A semantic \LaTeX{} macro & \texttt{\tbs JacobiP}, etc.\\
			\cline{2-4}
			\multicolumn{1}{l|}{} & Generic \LaTeX{} macro & All kinds of \LaTeX{} macros & \texttt{\tbs Rightarrow}, \texttt{\tbs alpha}, etc.\\
			\cline{2-4}
			\multicolumn{1}{l|}{} & Alphanumerical Expressions & Letters, numbers and general strings. & Depends on the order of symbols. $ab3$ is alphanumerical, while $4b$ are two nodes ($4$ and $b$).\\
			\cline{2-4}
			\multicolumn{1}{l|}{} & Symbols & All kind of symbols & `$@$', `$*$', `$+$', `$!$', etc.\\
			\hline
		\end{tabularx}
		\caption{A table of all kinds of nodes in a PoM syntax tree. Note that this table groups some types together for a better overview. For a complete list and a more detailed version see~\parencite{POM-Tagger}.}
		\label{tab:allTypesTable}
	\end{table} 
	
	\subsection{Problems with the Lookahead Approach}
	The lookahead functionality appears to solve the problems for functions. However, there is a problem with the lookahead functionality that \Cref{sec:problems} did not address. In some cases, the arguments of a function do not follow but precede the function node. 
	
	\begin{figure}
		\begin{minipage}{0.38\textwidth}
			\center
			\begin{tabular}{lc}
				\hline
				Notation & Expression \\
				\hline
				Infix & $(a+b) \cdot x$\\
				Prefix & $\cdot + a\ b\ x$\\
				Postfix & $a\ b + x\ \cdot$\\
				Functional & $\cdot(+(a, b), x)$\\
				\hline
			\end{tabular}
			\caption{The mathematical expression `$(a+b) \cdot x$' in infix, prefix, postfix and functional notation.}
			\label{tab:notations}
		\end{minipage}
	\end{figure}
	
	If we intently examine mathematical notations, we discover many different types of notations used to represent formulae. \Cref{tab:notations} illustrates the expression $(a+b)x$ in different notations. The \gls*{pol}\footnote{Also known as \textit{prefix notation}, \textit{Warsaw Notation} or \textit{{\L}ukasiewicz notation}. It was invented by J. {\L}ukasiewicz 1924 to create a parenthesis-free notation~\parencite{Hamblin1962}. Note that this notation is indeed parenthesis-free as long as all operators have the same arity.} (hereafter called prefix notation) places the operator to the left of/before its operands. The \gls*{rpol}\footnote{Also known as \textit{postfix notation}. Also invented by J. {\L}ukasiewicz. Same as \gls*{pol} it does not need parenthesis as long as all operators have the same arity.} (hereafter called postfix notation) does the opposite and places the operator to the right of/after its operands. The infix notation is commonly used in arithmetic and places the operator between its operands. This only makes sense if the operator is a binary operator.
	
	In mathematical expressions, notations are mostly mixed, depending on the case and number of operands. For example, infix notation is common for binary operators ($+$, $-$, $\cdot$, $\mod$, etc.), while functional notations are conveniently used for any kind of functions ($\sin$, $\cos$, etc.), and the postfix notation is often common for unary operators ($2!$, $-2$, etc.). Sometimes the same symbol is used in different notations to distinguish different meanings. For example, the `$-$' as a unary operator is used in prefix notation to indicate the negative value of its operand, such as in `$-2$'. Of course, `$-$' can also be the binary operator for subtraction, which is commonly used in infix notation. 
	
	Since it is more convenient to parse expressions using uniform notations, most programming 
	languages (and \gls*{cas} as well) internally use prefix or postfix notation and do not mix 
	the notations in one expression.
	The common practice in science is to use mixed notations in expressions. Since the \gls*{pom} has rarely implemented mathematical grammatical rules, it takes the input as it is and does not build an expression tree. Therefore, it parses all four examples from~\Cref{tab:notations} to four different \gls*{pom-pt}s rather than to one unique expression tree. In general, this is not a problem for our translation process since most \gls*{cas} are familiar with most common notations. Therefore, the translator does not need to know that `$a$' and `$b$' are the operands of the binary operator `$+$' in `$a+b$.' The translator could simply translate the symbols in `$a+b$' in the same order as they appear in the expression and the \gls*{cas} would understand it. However, there are two new problems with this approach.
	\begin{enumerate}\item \label{prob:1} The translated expression is only syntactically correct if the input expression was syntactically correct.
		\item \label{prob:2} We cannot translate expressions to \gls*{cas} which use non-standard notations.
	\end{enumerate}
	
	Problem~\ref{prob:1} should be obvious. Since we want to develop a translation tool and not a verification tool for mathematical \LaTeX{} expressions, we can assume syntactically correct input expressions and produce errors otherwise. Problem~\ref{prob:2} is more complex. If a user wants to support a \gls*{cas} that uses prefix or postfix notation by default, the translator would fail in its current state. Supporting \gls*{cas} with another notation would be a part of future work.
	
	Nonetheless, adopting different notations, in some situations, could also solve potential ambiguities. Consider the two potentially ambiguous examples in~\Cref{tab:amb_ex}. While a scientist would probably just ask for the right interpretation of the first example, \Maple{} automatically computes the first interpretation. On the other hand, \LaTeX{} automatically disambiguates the first example by only recognizing the very next element (single symbols or sequences in curly brackets) for the superscript and therefore displays the second interpretation. The second example should not be misinterpreted since this notation is the standard interpretation in science for the double factorial. We wrote the second interpretation with parentheses for pointing out that we mean the double factorial in this case. However, surprisingly, \Maple{} computes the first interpretation (the factorial of the factorial of $n$) again rather than the common standard interpretation.
	\begin{table}[ht]
		\centering
		\begin{tabular}{lccc}
			\hline
			& Text Format Expression & First Interpretation & Second Interpretation\\
			\hline
			1:~& \rule{0pt}{0.9\normalbaselineskip} $4\ \hat{\ }\ 2!$ & $4^{2!}$ & $4^2!$ \\
			2:~& $n!!$ & $(n!)!$ & $(n)!!$\\
			\hline
		\end{tabular}
		\caption{Potentially ambiguous examples using the factorial and double factorial symbols. One expression in a text format can potentially be interpreted in different ways.}
		\label{tab:amb_ex}
	\end{table}
	\vspace*{-0.5cm}
	
	In most cases, parentheses can be used to disambiguate expressions. We used them in~\Cref{tab:amb_ex} to clarify the different interpretations in Example 2. Note that the use of parentheses will not always resolve a mistaken computation. For example, there is no way to add parentheses to force \Maple{} to compute `$n!!$' as the double factorial function. Even `$(n)!!$' will be interpreted as `$(n!)!$'. Rather than using the exclamation mark in \Maple, one could also use the functional notation. For example, the interpretations `$(2!)!$' and `$(2)!!$' can be distinguished in \Maple{} by using \verb|factorial(factorial(2))| and \verb|doublefactorial(2)| respectively. We define the translations as follows:
	
	\begin{eqnarray*}
		\verb|n!| &\overset{\langMaple}{\mapsto}& \verb|factorial(n)|,\\
		\verb|n!!| &\overset{\langMaple}{\mapsto}& \verb|doublefactorial(n)|.
	\end{eqnarray*}

	Algorithm~\ref{alg:translation} does not allow this translation right now. It has no access to previously translated nodes in its current state. This problem is solved by the \gls*{teo} that stores and groups translated objects as lists. This allows one to access the latest translated expression and use it as the argument for the factorial function. Table~\ref{tab:teo-list} shows three examples for the \gls*{teo} list that groups some tokens.
	
	\begin{table}[ht]
		\centering
		\begin{tabular}{cc}
			\hline
			Input Expression & TEO List\\
			\hline
			$a+b$ & \verb|[a, +, b]|\\
			$(a+b)$ & \verb|[(a+b)]|\\
			$\frac{a}{b}-2$ & \verb|[(a)/(b), -, 2]|\\
			\hline
		\end{tabular}
		\caption{How the TEO-list groups subexpressions.}
		\label{tab:teo-list}
	\end{table}
	
	\subsection{Subtranslators}\label{sec:subtranslators}
	The \verb|SequenceTranslator| translates the \textit{sequence} and \textit{balanced expressions} in the \gls*{pom-pt}. If a node $n$ is a leaf and the represented symbol is an open bracket (parentheses, square brackets and so on), the following nodes are also taken as a \textit{sequence}. Combined with the recursive translation approach, the \verb|SequenceTranslator| also checks balancing of parentheses in expressions. An expression such as `$(a]$' produces a mismatched parentheses error. On the other hand, this is a problem for real interval expressions such as `$[a,b)$'. In the current version, the program cannot distinguish between mismatched parentheses and half-opened, half-closed intervals. Whether an expression is an interval or another expression is difficult to decide and can depend on the context. Also, the parentheses checker could simply be deactivated to allow mismatched parentheses in an expression. Another option is to use interval macros. e.g., \verb|\intcc@{a}{b} = [a,b]|.
	
	The \verb|SequenceTranslator| also handles positions of multiplication symbols. There are a couple of obvious choices to translate multiplication. The most common symbol for multiplications is still the white space (or no space between the tokens), as explained previously. Consider the simple expression `$2n\pi$'. The \gls*{pom-pt} generates a sequence node with three children, namely $2$, $n$ and $\pi$. This sequence should be interpreted as a multiplication of the three elements. The \verb|SequenceTranslator| checks the types of the current and next nodes in the tree to decide if it should add a multiplication symbol or not. For example, if the current or next node is an operator, a relation symbol or an ellipsis, there will be no multiplication symbol added. However, this approach implies an important property. The translator interprets all sequences of nodes as multiplications as long as it is not defined otherwise. This potentially produces strange effects. Consider an expression such as `$f(x)$'. Translating this to \Maple{} will give \verb|f*(x)|. But we do not consider this translation to be wrong, because there is a semantic macro to represent functions. In this case, the user should use \verb|\f{f}@{x}| instead of \verb|f(x)| to distinguish between $f$ as a function call and $f$ as a symbol.
	
	\begin{algorithm}[!ht]
		\caption{The translate function of the MacroTranslator. This code ignores error handling.}\label{alg:macro-translation}
		\begin{algorithmic}[1]
			\Require 
			\Statex \textit{macro} - node of the semantic macro. 
			\Statex \textit{args} - list of the following siblings of \textit{macro}. 
			\Statex \textit{lexicon} - lexicon file
			\Ensure 
			\Statex Translated semantic macro.
			
			\Procedure{translate\_macro}{\textit{macro}, \textit{args}, \textit{lexicon}}
			\State \textit{info} = \textit{lexicon}.getInfo(\textit{macro});\label{line:get_info}
			\State \textit{argList} = new List(); \Comment{create a sorted list for the translated arguments.}
			\State \textit{next} = \textit{args}.getNextElement();
			\If{\textit{next} is caret}\label{line:next_caret}
			\State \textit{power} = translateCaret(\textit{next});
			\State \textit{next} = \textit{args}.getNextElement();
			\EndIf
			
			\While{\textit{next} is $[$}\label{line:next_optional} \Comment{square brackets starts a balanced sequence of optional arguments.}
			\State \textit{optional} = {\scriptsize TRANSLATE\_UNTIL\_CLOSED\_BRACKET}(\textit{args});
			\State \textit{argList}.add(\textit{optional});
			\State \textit{next} = \textit{args}.getNextElement();
			\EndWhile
			
			\State \textit{argList}.add( {\scriptsize TRANSLATE\_PARAMETERS}(\textit{args}, \textit{info}) ); \label{line:trans_paras} \Comment{number is given in \textit{info}.}
			\State {\scriptsize SKIP\_AT\_SIGNS}( \textit{args}, \textit{info} );  \label{line:skip_ats} \Comment{number is given in \textit{info}.}
			\State \textit{argList}.add( {\scriptsize TRANSLATE\_VARIABLES}(\textit{args}, \textit{info}) ); \label{line:trans_vars} \Comment{number is given in \textit{info}.}
			
			\State \textit{pattern} = \textit{info}.getTranslationPattern();
			\State \textit{translatedMacro} = \textit{pattern}.fillPlaceHolders(\textit{argList});\label{line:fill_pattern}
			\If{\textit{power} is not \NULL}\label{line:shifted_exp}
			\State \textit{translatedMacro}.add(\textit{power});
			\EndIf
			\State \Return\ \textit{translatedMacro};
			\EndProcedure
		\end{algorithmic}
	\end{algorithm}
	
	The translation process for the \Macro s is complex, so there is a special class, the \verb|MacroTranslator|, that handles those nodes in the \gls*{pom-pt}. Algorithm~\ref{alg:macro-translation} explains the \verb|MacroTranslator| without error handling. It has extracted necessary information from the \gls*{pom-pt}, such as how many arguments this function has, in Line~\ref{line:get_info}. It also processes the following siblings to translate the arguments. The \verb|MacroTranslator| will be called in Line~\ref{line:transLeaf} in Algorithm~\ref{alg:translation}, since the macro is a leaf node in the \gls*{pom-pt}. The following cases describe the different kinds of the following siblings after a semantic macro node. Those can be:
	\begin{itemize}
		\item an exponent, such as for `\verb|^2|' right after the macro node (Line~\ref{line:next_caret});
		\item an optional parameter in square brackets right after the macro node or after an exponent (Line~\ref{line:next_optional});
		\item a parameter in curly brackets (a \textit{sequence} node in the \gls*{pom-pt}) if none of the above and no `$@$' symbols were passed (Line~\ref{line:trans_paras});
		\item `$@$' symbols (Line~\ref{line:skip_ats}); or
		\item a variable in curly brackets (a \textit{sequence} node) after the `$@$' symbols were 
		passed (Line~\ref{line:trans_vars}).
	\end{itemize}
	
	All cases before the `$@$' symbols are optional. The \verb|MacroTranslator| removes all following siblings according to the number of expected parameters and variables. Parameter and variable nodes are translated separately. If an exponent was registered right after the semantic macro node, it will be shifted to the end in Line~\ref{line:shifted_exp}. The macro itself will be translated by putting all translated parameters and variables into the translation pattern (Line~\ref{line:fill_pattern}).
	
	Following siblings after the macro was translated (with all arguments) do not belong to the semantic macro. If the next node is an exponent, the translated macro is the base. Table~\ref{tab:multi-expo} shows an example for the translation of the trigonometric cosine function with multiple exponents.
	
	\begin{table}[ht]
		\centering
		\begin{tabular}{lccc}
			& Semantic \LaTeX & & \Maple{} \\
			\hline
			Text Representation & \rule{0pt}{1.0\normalbaselineskip} \verb|\cos^n@{x}^m| & $\overset{\langMaple}{\mapsto}$ & \verb|((cos(x))^(n))^m| \\
			Displayed As & \rule{0pt}{1.0\normalbaselineskip} $\cos^n(x)^m$ & & $\left( \cos(x)^n \right)^m$ \\
			\hline
		\end{tabular}
		\caption{A trigonometric cosine function example with exponents before and after the argument.}
		\label{tab:multi-expo}
	\end{table}
	
	\section{\Maple{} to Semantic \LaTeX{} Translator}\label{sec:backward-translation}
	In this section, we will discuss several techniques to access the parse tree of \Maple's input. The translation process from this parse tree then follows the same principle as for the forward translations. Instead of writing a custom \Maple{} syntax parser, we use \Maple's internal data structure to obtain the syntax tree of the input\footnote{A license of \Maple{} is mandatory to perform backward translations. Our translator uses the version \Maple{} 2016.}. \Maple{} allows several different input styles. The \texttt{1D} input is mainly used for programming purposes 
	and is also used to perform our translations. Internally, \Maple{} uses a \gls*{dag} for syntax trees.
	
	\begin{wrapfigure}[7]{r}{0.61\textwidth}
		\vspace{-15pt}
		\centering
		\includegraphics[clip, trim=0.5cm 0.5cm 0.5cm 0.5cm, scale=0.5]{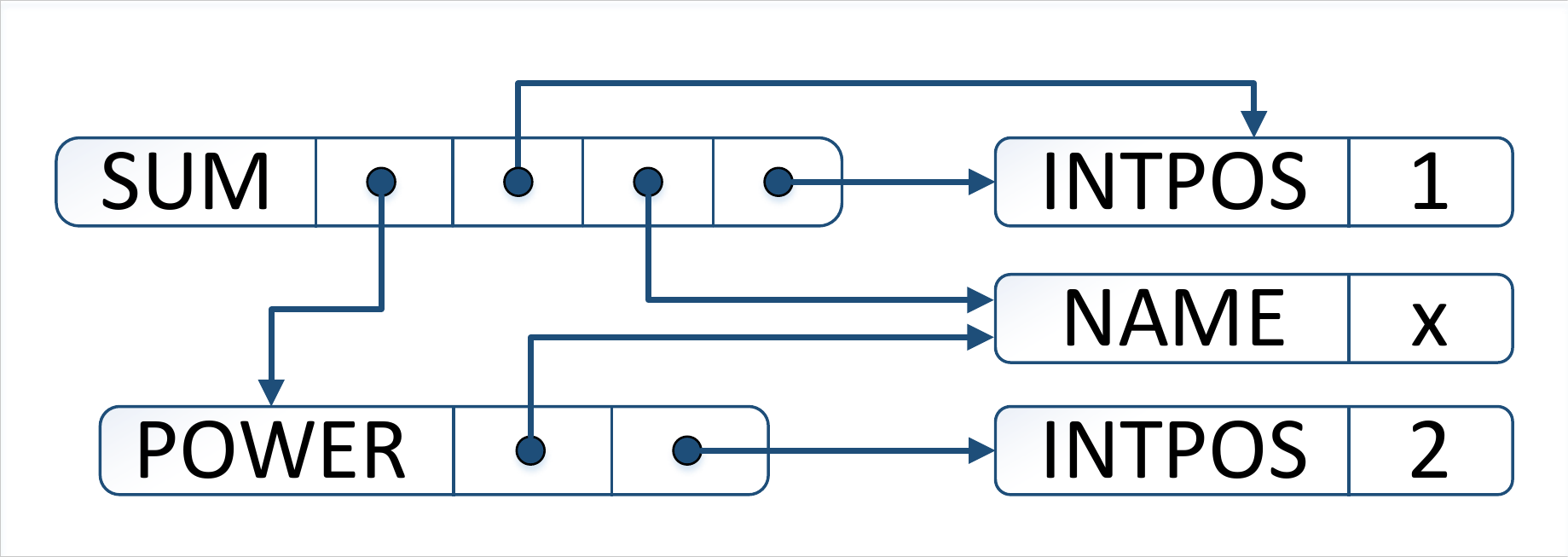}
		\vspace{-5pt}
		\caption{The internal \Maple{} DAG representation of $x^2+x$.}
		\label{fig:internal-maple-dag}
	\end{wrapfigure}
	
	Each node in the \gls*{dag} stores its children and has a header which defines the type and the length of the node. Consider the polynomial $x^2+x$. Figure~\ref{fig:internal-maple-dag} illustrates the internal \gls*{dag} representation with headers and arguments.
	
	One can access the internal data structure of expressions via the \texttt{ToInert} command, which returns the \inertF. The \inertF{} format is a nested list\footnote{The nested list is a tree representation of a \gls*{dag} that splits nodes with multiple parents into multiple nodes so that each node has only one parent node.} of the internal \gls*{dag} for the given expression. Some of the important types for the nodes are specified in \Cref{tab:maple-types}. The translator uses the \texttt{OpenMaple}~\parencite[\S 14.3]{MAPLE:ProgrammingGuide} \gls*{api} for interacting with \Maple's kernel implementation.
	
	\begin{table}[ht]
		\centering
		\begin{tabular}{cp{10cm}}
			\hline
			Type & Explanation\\
			\hline
			SUM & Sums. Internally stored with factors for each summand, i.e., `$x+y$' would be stored as `$x \cdot 1 + y \cdot 1$'.\\
			PROD & Products.\\
			EXPSEQ & Expression sequence is a kind of list. The arguments of functions are stored in such sequences.\\
			INTPOS & Positive integers.\\
			INTNEG & Negative integers.\\
			COMPLEX & Complex numbers with real and imaginary part.\\
			FLOAT & Float numbers are stored in the scientific notation with integer values for the exponent $n$ and the significand $m$ in $m \cdot 10^n$.\\
			RATIONAL & Rational numbers are fractions stored in integer values for the numerator and positive integers for the denominator.\\
			POWER & Exponentiation with expressions as base and exponent.\\
			FUNCTION & Function invocation with the name, arguments and attributes of the function.\\
			\hline
		\end{tabular}
		\caption{A subset of important internal \Maple{} data types. See~\parencite{MAPLE:ProgrammingGuide} for a complete list.}
		\label{tab:maple-types}
	\end{table}
	
	\subsection{Automatic Changes of Inputs in Maple}\label{subsec:maple-probs}
	\Maple{} evaluates inputs automatically and changes the input into an internal representation. This internal representation may differ to the input. One example has already been given with \Cref{fig:internal-maple-dag}, where each summand of a sum is stored with a factor. Here is a list of all internal changes that occur for inputs.
	
	\begin{itemize}
		\item \Maple{} evaluates input expressions immediately.
		\item There is no data type to represent square roots such as $\sqrt{x}$ (or $n$-th roots). Therefore, \Maple{} stores roots as an exponentiation with a fractional exponent. For example, $\sqrt{x}$ is stored as $x^{\frac{1}{2}}$.
		\item There is no data type for subtractions, only for sums. Negative terms are changed to absolute values times `$-1$'. For example, $x-y$ is stored as $x + y \cdot (-1)$. 
		\item Floating point numbers are stored using scientific notation with a mantissa and an exponent in the base $10$. For example, $3.1$ is internally represented as $31 \cdot 10^{-1}$.
		\item There is only a data type for rational numbers (fractions with an integer numerator and a positive integer denominator), but not for general fractions, such as $\frac{x+y}{z}$. This will be automatically changed to $(x+y)\cdot z^{-1}$.
	\end{itemize}
	
	There are unevaluation quotes implemented to avoid evaluations on input expressions. Table~\ref{tab:unevaluation-quotes} gives an example how unevaluation quotes work.
	
	\begin{table}[ht]
		\centering
		\begin{tabular}{lcc}
			\hline
			& Without unevaluation quotes & With unevalation quotes\\
			\hline
			Input expression:~& \texttt{sin(Pi)+2-1} & \texttt{'sin(Pi)+2-1'}\\
			Stored expression:~& 1 & \texttt{sin(Pi)+1}\\
			\hline
		\end{tabular}
		\caption{Example of unevaluation quotes for \texttt{1D} \Maple{} input expressions.}
		\label{tab:unevaluation-quotes}
	\end{table}
	
	Since we want to keep a translated expression similar to the input expression, we implemented some cosmetic rules 
	for backward translations which solve or reduce the effects due to the list of changes above. 
	\begin{itemize}
		\item We use unevaluation quotes to suppress evaluations of the input.
		\item We perform a reordering of factors and summands so that negative factors appear in front of the summand. This gives us the opportunity to translate $x - y$ to $x - y$ instead of $x + y \cdot (-1)$.
		\item We introduced new internal data types \texttt{MYFLOAT} and \texttt{DIVIDE} to translate floats and fractions in more convenient notations.
	\end{itemize}
	
\begin{figure}[t!]
		\centering
		\includegraphics[clip, trim=0.1cm 0.1cm 0.1cm 0.1cm, scale=0.7]{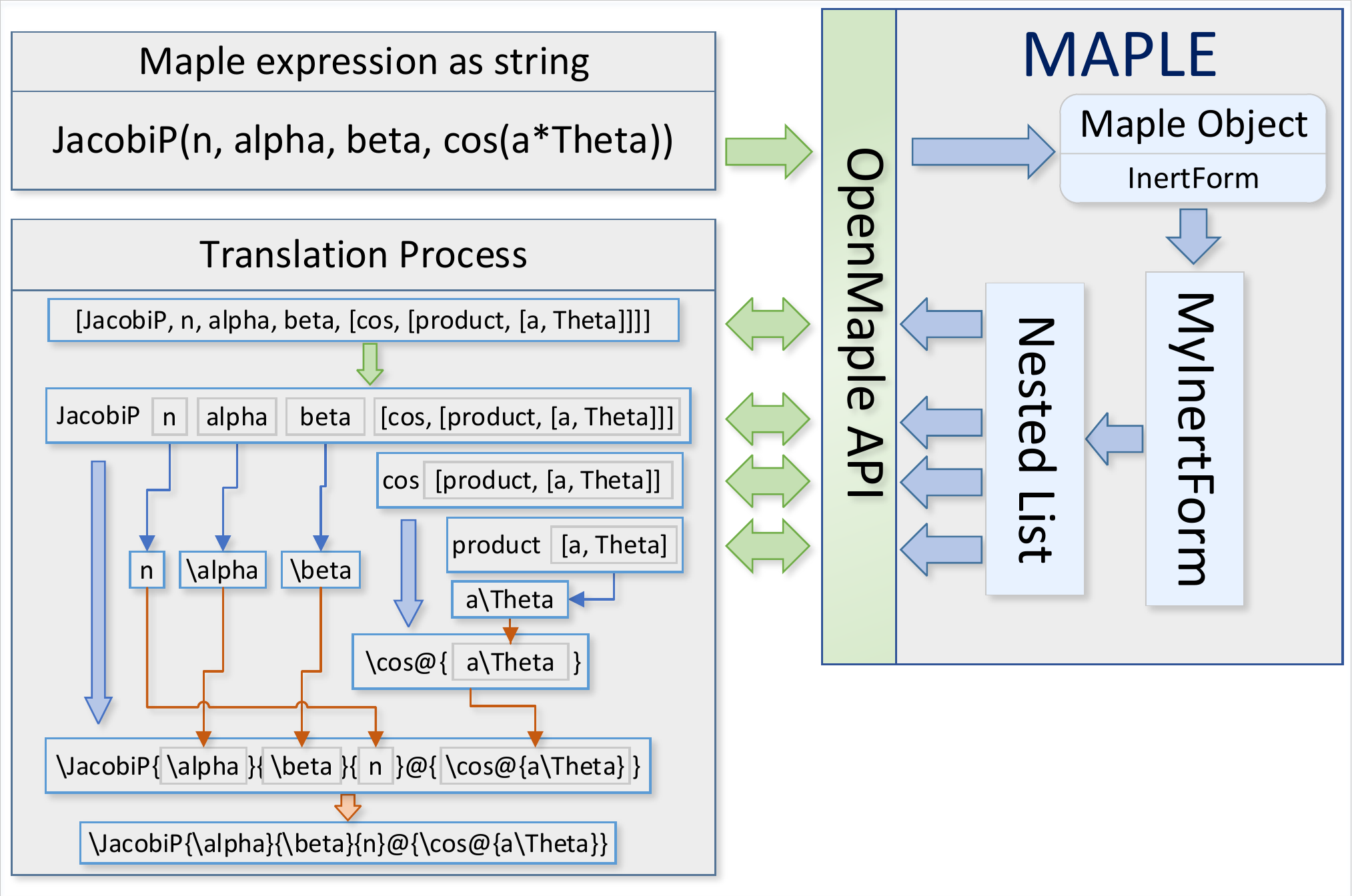}
		\caption{A scheme of the backward translation process from \Maple{} for the Jacobi polynomial expression $P_n^{(\alpha , \beta)}(\cos(a\Theta))$. The input string is converted by the \Maple{} kernel into the nested list representation. This list is translated by subtranslators (blue and red arrows). A function translation (bold blue arrows) is again realized using translation patterns to define the position of the arguments (red arrows).}
		\label{fig:backward-trans}
	\end{figure}

	The translation process then follows the same principle as for the forward translations. Since the syntax tree of \Maple{} is an expression tree, we do not need to implement special reordering or grouping algorithms to perform backward translations. Translations for functions are also realized via patterns and placeholders. Figure~\ref{fig:backward-trans} illustrates the backward translation process for the Jacobi polynomial example from \Cref{tab:JacobiP-usecase}.
	
	\section{Evaluation}\label{sec:evaluation}
	We implemented three approaches to evaluate whether a translation was \textit{appropriate} or \textit{inappropriate}.
	\begin{enumerate}
		\item \textbf{Round Trip Tests}:~translates expressions back and forth and analyzes the changes.
		\item\label{approach-2} \textbf{Function Relation Tests (Symbolical)}:~translates mathematically proven equivalent expressions from one system to a \gls*{cas} and evaluates whether the relation remains valid via symbolical equivalence checks.
		\item \textbf{Numerical Tests}:~takes the same equations from Approach~\ref{approach-2} but evaluates them on specific numerical values to test whether the translation was {\it appropriate}.
	\end{enumerate}
	
	\subsection{Round Trip Tests}\label{sec:round-trip}
	A round trip test always starts with a valid expression either in semantic \LaTeX{} or in \Maple. A translation from one system to another is called a \textbf{step}. A complete round trip translation (two steps) is called \textbf{one cycle}. A \textbf{fixed point representation} (or short fixed point) in a round trip translation process is a string representation that is identical to all string representations in the following cycles. Table~\ref{tab:fixpoint} illustrates an example of a round trip test which reaches a fixed point for the mathematical expression
	\begin{equation}
	\frac{\cos\left(a\Theta\right)}{2}.
	\end{equation}
	
	\begin{table}[ht]
		\centering
		\begin{tabular}{cc}
			\hline 
			Steps & semantic \LaTeX{}/\Maple{} representations\\
			\hline
			\rule{0pt}{0.9\normalbaselineskip}0 & \verb|\frac{\cos@{a\Theta}}{2}|\\
			1 & \verb|(cos(a*Theta))/(2)| \\
			2 & \verb|\frac{1}{2}\idot\cos@{a\idot\Theta}| \\
			3 & \verb|(1)/(2)*cos(a*Theta)|\\
			4 & \verb|\frac{1}{2}\idot\cos@{a\idot\Theta}| \\
			\hline
		\end{tabular}
		\caption{A round trip test reaching a fixed point.}
		\label{tab:fixpoint}
	\end{table}
	
	Step 4 is identical to step 2, and since the translator is a deterministic algorithm, it can be easily shown that step 2 and step 3 are fixed-point representations for semantic \LaTeX{} and \Maple.
	
	There is currently only one exception known where a round trip test does not reach a fixed point representation:~Legendre's incomplete elliptic integrals~\parencite[(19.2.4-7)]{NIST:DLMF} are defined with the amplitude $\phi$ in the first argument in the \DLMF, while \Maple{} takes the trigonometric sine of the amplitude as the first argument. Therefore, the forward and backward translations are defined as
	\begin{eqnarray}
	\verb|\EllIntF@{\phi}{k}| & \overset{\langMaple}{\mapsto} & \verb|EllipticF(sin(phi),k)|,\\
	\verb|\EllIntF@{\asin@{\phi}}{k}| & \overset{\langMaple}{\mapsfrom} & \verb|EllipticF(phi,k)|,
	\end{eqnarray}
	and the round-trip translations produce infinite chains of sine and inverse sine calls because there are no evaluations involved. 
	
	The round trip tests are very successful, but they only detect errors in string representations. However, because of the simplification techniques of fixed points, we are able to at least detect logical errors in one system:~\Maple. On the other hand, these tests cannot determine logical errors in the translations between the two systems. Suppose we mistakenly defined an \textit{inappropriate} forward and backward translation for the sine function
	\begin{eqnarray}
	\verb|\sin@{\phi}| & \overset{\langMaple}{\leftrightarrow} & \verb|cos(phi)|,\label{eq:wrong-trans-1}\\
	\verb|\cos@{\phi}| & \overset{\langMaple}{\leftrightarrow} & \verb|sin(phi)|.\label{eq:wrong-trans-2}
	\end{eqnarray}
	In that case the round trip test would not detect any errors but reaches a fixed point representation.
	
	\subsection{Function Relation Tests}\label{sec:relation-tests}
	The \gls*{dlmf} is a compendium for special functions and orthogonal polynomials and lists many relations between the functions and polynomials. The idea of this evaluation approach is to translate an entire relation and test whether the relation remains valid after performing the translations.
	
	With this idea, we can detect {\it inappropriate} translations such as in \eqref{eq:wrong-trans-1} and \eqref{eq:wrong-trans-2}. Consider the \gls*{dlmf} equation for the sine and cosine function~\parencite[(4.21.2)]{NIST:DLMF}
	\begin{equation}
	\sin \left(u+v\right) = \sin{u}\cos{v} + \cos{u}\sin{v}.
	\end{equation}
	Assume the translator would forward translate the expression based on (\ref{eq:wrong-trans-1},~\ref{eq:wrong-trans-2}). Then
	\begin{eqnarray}
	\verb|\sin@{u + v}| & \overset{\langMaple}{\mapsto} & \verb|cos(u + v)|,\\
	\verb|\sin@@{u}\cos@@{v}| & \overset{\langMaple}{\mapsto} & \verb|cos(u)*sin(v)|,\\
	\verb|\cos@@{u}\sin@@{v}| & \overset{\langMaple}{\mapsto} & \verb|sin(u)*cos(v)|.
	\end{eqnarray}
	This produces the equation in \Maple
	\begin{equation}
	\cos\left(u+v\right) = \cos{u}\sin{v} + \sin{u}\cos{v},
	\end{equation}
	which is wrong. Since the expression is correct before the translation, we conclude that there was an error 
	during the translation process and our defined translations were {\it inappropriate}.
	
	There are two essential problems with this approach. Testing whether expressions are {\it appropriate} representations of each other is a challenging task for \gls*{cas} and they often have difficulties testing simple equations symbolically. For example, 
	consider
	\parencite[(4.35.34)]{NIST:DLMF}
	\begin{equation*}
		\sinh \left( x+\iunit y \right) = \sinh x \cos y + \iunit \cosh x \sin y,
	\end{equation*}
	as a difference of the left- and right-hand sides cannot be simplified to zero by default. Furthermore, this approach only checks forward translations because there is no way to automatically check whether two \LaTeX{} expressions are {\it appropriate} or {\it inappropriate} representations of each other (again this could become feasible with our translator). We use \Maple's \texttt{simplify} function to check if the difference of the left-hand side and the right-hand side of the equation is equal to zero. In addition, we use \texttt{simplify} and check if the division of the right-hand side by the left-hand side returns a numerical value or not. The simplification function is the most powerful function to check whether two expressions are {\it appropriate} representations in \Maple. However, there are several cases where simplification fails. Because of implementation details, there are some techniques that help \Maple{} to find possible simplifications. For example, we can force \Maple{} to convert the formula
	\begin{equation}
	\sinh{x} + \sin{x}
	\end{equation}
	to an {\it appropriate} representation using their exponential representations, namely
	\begin{equation}
	\frac{1}{2}\expe^x - \frac{1}{2}\expe^{-x} - \frac{1}{2} \iunit \left( \expe^{\iunit x}-\expe^{-\iunit x} \right).
	\end{equation}
	With such pre-conversions, we are able to improve the simplification process in \Maple. However, the limitations of the \texttt{simplify} function are still the weakest part of this verification approach. Consider the complex example~\parencite[(12.7.10)]{NIST:DLMF}
	\begin{equation}\label{eq:branch-cut-near}
	\displaystyle U(0,z) = \sqrt{\frac{z}{2\cpi}} K_{\frac{1}{4}}\left(\frac{1}{4}z^2\right),
	\end{equation}
	where $U(0,z)$ is the parabolic cylinder function and $K_\nu(z)$ is the modified Bessel function of the second kind. Both functions are well-defined in both systems and we can define a \textit{direct} translation for~\eqref{eq:branch-cut-near}. 
	The modified Bessel function of the second kind has its branch cut in \Maple{} and in the \gls*{dlmf} at $z < 0$. However, the argument of $K$ contains a $z^2$. If $|\ph{z}| \in \left(\frac{\cpi}{2}, \cpi\right)$ the value of the right-hand side of~\eqref{eq:branch-cut-near} would be no longer on the principal branch. \Maple{} will still compute the principal values independently of the value of $z$ and the translation
	\begin{equation}
	\verb|\BesselK{\frac{1}{4}}@{\frac{1}{4}z^2}| \overset{\langMaple}{\mapsto} \verb|BesselK(1/4,(1/4)*z^2)|
	\end{equation}
	is {\it inappropriate} if $|\ph{z}| \in \left(\frac{\cpi}{2}, \cpi\right)$. One should instead use the analytic continuation for the right-hand side of (\ref{eq:branch-cut-near}). 
	To evaluate such complex cases, the previous checks for {\it appropriate} representations in \gls*{cas} are insufficient. Therefore we implement numerical tests as an additional step.
	
	\subsection{Numerical Tests}\label{sec:numerical-tests}
	Consider the difference of the left- and right-hand sides of equation~\eqref{eq:branch-cut-near}, namely
	\begin{equation}\label{eq:difference}
	D(z) := U(0,z) - \sqrt{\frac{z}{2\cpi}} K_{\frac{1}{4}}\left(\frac{1}{4}z^2\right).
	\end{equation}
	Table~\ref{tab:computations-for-difference} presents four numerical evaluations for $D(z)$, one value for each quadrant in the complex plane.
	\begin{table}[ht]
		\centering
		\begin{tabular}{rcc}
			\hline
			$z\ \ $ & & $D(z)$\\
			\hline
			\tableRowSpace{} $1+\iunit$ & & $2 \cdot 10^{-10} - 2 \cdot 10^{-10} \iunit$\\
			$-1+\iunit$& & $2.222121916 - 1.116719816 \iunit$\\
			$-1-\iunit$& & $2.222121916 + 1.116719816 \iunit$\\
			$1-\iunit$ & & $2 \cdot 10^{-10} + 2 \cdot 10^{-10} \iunit$\\
			\hline
		\end{tabular}
		\caption{Four numerical evaluations of $D(z)$ in \Maple.}
		\label{tab:computations-for-difference}
	\end{table}
	
	Considering machine accuracy and the default precision to $10$ significant digits, we can regard the first and last values as zero differences. While this evaluation is very powerful, it has a significant problem. Even when all tested values return zero, it does not prove that~\eqref{eq:branch-cut-near} was {\it appropriately} translated. When the values are different from zero, it does indicate that there might be an error satisfying one of the four cases~\parencite{NumericalTests:Paper}:
	\begin{enumerate}
		\item the numerical engine tests invalid combinations of values;
		\item the translation was {\it inappropriately} defined;
		\item there may be an error in the \gls*{dlmf} source; or
		\item there may be an error in \Maple.
	\end{enumerate}
	
	\subsection{Results}\label{sec:test-summary}
	There are currently 685 \Macro{}s\footnote{The DLMF/DRMF semantic macros are still a work in progress, and the total number is constantly changing.} in total, and 665 of them were implemented in the translator engine. We defined forward translations to \Maple{} for 201 of the macros and backward translations from \Maple{} for 195 functions. 
	
	The \gls*{dlmf} provides a dataset of \LaTeX{} expressions with semantic macros. We extracted 4087 equations from the \gls*{dlmf} and applied our round-trip and relation tests on them. The translator was able to translate 2405\footnote{All percentages are approximately calculated.} (58.8\%) of the extracted equations without errors. 
	Simplification techniques of \Maple{} were successfully verified for 660 (27.4\%) of the translated expressions. 
	We applied additional numerical tests for the remaining 1745 equations. For 418 (24\%) of them, the numerical tests were valid. More detailed results for numerical and symbolical tests were presented in~\parencite{NumericalTests:Paper}.

	The evaluation techniques have proven to be very powerful for evaluating \gls*{cas} and online mathematical compendia such as the \gls*{dlmf}. During the evaluations, we were able to detect several errors in the translation and evaluation engine, and also discovered two errors in the \gls*{dlmf} and one error in \Maple's \texttt{simplify} function.
	
	The numerical test engine was able to discover a sign error in equation~\parencite[(14.5.14)]{NIST:DLMF}\footnote{The equation had originally been stated as shown in (\ref{eq:signerror}). The error was reported on 10th April 2017.}
	\begin{equation}\label{eq:signerror}
	\displaystyle \mathsf{Q}^{-1/2}_{\nu}\left(\cos\theta\right)=-\left(\frac{\pi}{2\sin\theta}\right)^{1/2}\frac{\cos\left(\left(\nu+\frac{1}{2}\right)\theta\right)}{\nu+\frac{1}{2}}.
	\end{equation}
	The error can be found on~\parencite[p. 359]{NIST:Handbook} and has been fixed in the \gls*{dlmf} with version 1.0.16. The same engine also identified a missing comma in the constraint of~\parencite[(10.16.7)]{NIST:DLMF}. The original constraint was given by $2\nu \neq -1, -2 -3, \ldots$, with a missing comma after the $-2$.
	
	We have also noticed that our testing procedure is able to identify errors in CAS procedures,
	namely the \Maple{} {\tt simplify} procedure. The left-hand side of~\parencite[(7.18.4)]{NIST:DLMF} is given by 
	\begin{equation*}
		\frac{{\mathrm{d}}^{n}}{{\mathrm{d}z}^{n}}\left(e^{z^{2}}\operatorname{erfc}z\right), \quad n = {0,1,2,\ldots},
	\end{equation*}
	where $e$ is the base of the natural logarithm, and $\operatorname{erfc}$ is the
	complementary error function. Our translation correctly produces 
	\begin{equation*}
		\verb|diff((exp((z)^(2))*erfc(z)), [z$(n)])|.
	\end{equation*}
	However, the \Maple{ 2016} \verb|simplify| function falsely returns $0$ for the translated left-hand side.
	Maplesoft has confirmed in a private communication that this is indeed a defect in \Maple{ 2016}.
	Furthermore, although the nature of the defect changes, the defect still persists in \Maple{ 2018}
	as of the publication of this manuscript.
	
	\section{Conclusion \& Future Work}\label{ch:conc-future-work}
	During this project we uncovered several problems that needs to be solved before providing a translation of mathematical expressions between two systems. The translator concept has proven itself by discovering errors in the online \gls*{dlmf} compendia and the test cases have also shown how difficult it is to validate translated expression. Our validation techniques also assume the correctness of simplification and computational algorithms in \gls*{cas}. However, combining those techniques and automatically running translation checks, not only can discover errors in mathematical compendia but can also detect errors in simplifications or computations of the \gls*{cas}.
	
	The tasks for future work are diverse. The main task is to improve the translator by implementing more functions and features. For example, for the current state, only translations to \Maple's standard function library were implemented. \Maple{} allows one to load extra packages dynamically and therefore support an enhanced set of functions. This feature would drastically increase the number of possible translations. With such improvements, further work on evaluation techniques become worthwhile to evaluate the \gls*{dlmf} and \gls*{cas}.
	Increasing the amount of translatable formulae in the \gls*{dlmf} and improving the verificiation 
	techniques are also parts of ongoing projects. 
	
	The translator was designed to be easily extendable. This allows one to implement translations for other \gls*{cas} without much effort.
	However, most \LaTeX{} sources, such as in arXiv, are given in generic \LaTeX{}. Semantic \LaTeX{}, which is a prerequsite for our translator, is currently prevalent in the DLMF and DRMF projects alone. Therefore, without exclusively given semantic information, the translator is not able to translate functions. Currently, we are working on mathematical information retrieval techniques which will allow for an extension of the translator to generic \LaTeX{} inputs.
	
	Further improvements for numerical tests could be to perform tests for specific (critical) values \parencite{DBLP:journals/aaecc/BeaumontBDP07} with respect to the involved functions. Beamont and collaborators tested identities for multivalued elementary functions by choosing sample points from regions with respect to branch cuts for functions. Choosing sample points from those regions could significantly improve the success rate of the numerical tests.
	
	\noindent{\bf\large Acknowledgements} This work was supported by the German Research Foundation (DFG grant GI-1259-1).
	
	\emergencystretch=1em
	\printbibliography
	
\end{document}